\documentclass[]{interact}

\usepackage{epstopdf}
\usepackage{subfigure}

\usepackage[numbers,sort&compress,merge]{natbib}
\bibpunct[, ]{(}{)}{,}{n}{,}{,}

\theoremstyle{plain}

\theoremstyle{definition}

\theoremstyle{remark}

\usepackage[utf8]{inputenc}
\usepackage[frenchb]{babel}
\usepackage{amsmath}
\usepackage{amsfonts}
\usepackage{amssymb}
\usepackage{fancyhdr,xcolor,soul}
\usepackage{textcomp} 

\begin{document}
\title{Symmetry breaking in linear multipole traps}

\author{\name{J. Pedregosa-Gutierrez\textsuperscript{a}\thanks{CONTACT J. Pedregosa-Gutierrez. Email: jofre.pedregosa@univ-amu.fr}, C. Champenois\textsuperscript{a}, M.R. Kamsap\textsuperscript{a,b}, G. Hagel\textsuperscript{a}, M. Houssin\textsuperscript{a}, and M. Knoop\textsuperscript{a}}
\affil{\textsuperscript{a}Aix-Marseille Universit\'e, CNRS, PIIM, UMR 7345, 13397 Marseille, France; \textsuperscript{b}Département de Physique, Université des Sciences et Techniques de Masuku, BP 943, Franceville, Gabon}}
\articletype{ARTICLE TEMPLATE}

\maketitle

\date{\today}

\begin{abstract}
Radiofrequency multipole traps have been used for some decades in cold collision experiments, and are gaining interest for precision spectroscopy due to their low micromotion contribution, and the predicted unusual cold-ion structures. However, the experimental realisation is not yet fully controlled, and open questions in the operation of these devices remain. We present experimental observations of symmetry breaking of the trapping potential in a macroscopic octupole trap with laser-cooled ions. Numerical simulations have been performed in order to explain the appearance of additional local potential minima, and be able to control them in a next step.  We characterize these additional potential minima, in particular with respect to their position, their potential depth and their probability of population as a function of the radial and angular displacement of the trapping rods. 
\end{abstract}


\section{Introduction}
The linear radio-frequency (RF) ion trap is a device which is widely used in physics and chemistry, and its quadrupole version can be found in experiments ranging from quantum computing~\cite{blatt08}, metrology~\cite{chen17}, and the observation of time-crystals \cite{zhang17}, to the study of large ion crystals \cite{hornekaer02,kamsap15a}. Compared to the harmonic potential of the quadrupole configuration, higher order traps generate a flatter electric field in the centre of the trap, which induces less RF-driven motion and therefore, less RF-heating than quadrupole traps~\cite{gerlich92}. For this reason multipole traps have been traditionally used in the cold collision community~\cite{asvany13} at temperatures of a few $K$, reached by buffer gas cooling. Compared to the quadrupole trap, the higher-order radial potential results in a different ion density distribution \cite{champenois09}, the difference being more pronounced when the trapped ions are laser-cooled. Laser-cooled ions in multipole traps are expected to lead to Coulomb crystals of a new kind~\cite{okada07,champenois10,marciante12}, in particular ion rings, as  has been shown numerically. \\

Compared to an ion string, the ion ring has a higher degree of symmetry and identical inter-ion distance. This system could therefore be an ideal candidate to study Hawking radiation from an acoustic black hole~\cite{horstmann10}, the Aharonov-Bohm effect~\cite{noguchi14}, or the possibility to generate space-time crystals~\cite{li12}.
Ion rings in ion traps have also been proposed for clock applications~\cite{champenois10}, to study second order phase transitions~\cite{cartarius13}, or discrete kink solitons~\cite{mielenz13,landa14}, and to generate Rydberg rings as a source for the generation of non-classical light~\cite{olmos10}.

The first report of ion crystals in a multipole trap \cite{okada07} leaves room for interpretation of the images, and has not been reproduced. Recently, ion rings have been created experimentally in the 2D versions of the quadrupole storage ring \cite{waki92} (a linear quadrupole trap closed in itself). Micro-fabricated surface traps with a large number of electrodes  allow to trap up to 400 ions in a ring structure with a radius of 624~$\mu m$ and an average separation of 9~$\mu m$ \cite{tabakov15}, or up to 15~ions in a 90~$\mu$m- ring \cite{li17}.

Multipole traps have shown unexpected features, in particular with cold ions. Some years ago, Otto et al. \cite{otto09} have reported the existence of local minima in the trapping potential of a 22-pole trap, as evidenced by the spatial distribution of H$^{-}$ ions at a temperature of 170~K. The same group has also reported incomplete cooling of trapped ions \cite{endres17}. Recently, numerical simulations in the case of a non-ideal 3D octupole and a 22 pole trap have reported potential minima with a local depth in the $meV$-range \cite{fanghanel2017}.

Our experimental set-up is a double linear RF trap composed of a quadrupole and an octupole trap mounted along the same $z$-axis, in order to allow shuttling of the ion clouds between the various trapping zones \cite{champenois13bis, kamsap15t}.
The present work is focused on the octupole trap, where we have recently observed the evidence of three local potential minima which do not contain an identical number of ions. Motivated by this observation we have characterised the behaviour of the ions trapped therein and realised a numerical simulation reproducing all trap parameters in order to identify the causes of such ion distribution. It turns out that the observed minima are created by extremely small deviations from the ideal trap geometry. In order to control and correct the observed non-ideal potential we have carried out a systematic study of the created additional potential minima as a function of the radial and angular displacements of the trap rods. This analysis results in an estimation of the geometric defects and of the local potential depth, and therefore a limit to the ion temperature, which is in good agreement with our observations. The ultimate goal of this study is to serve as a starting point for the correction of the asymmetry of the multipole potential.


This article is organized as follows. First, the analytic description of the potential inside the multipole trap is reviewed. We then report our recent experimental work concerning laser-cooled ions in an octupole trap. In the following section, the numerical study of the effect of a single rod displacement on the generation of local minima independently of the trap dimensions is presented. We then identify the characteristics of the defects responsible for our experimental observations and numerically reproduce  the relative number of ions in each potential minimum.

\section{Trapping potentials in a linear multipole RF trap}
The radial potential generated inside an ideal and infinitely long linear RF trap formed by $2k$ rods alternately polarised by a time oscillating electric potential \cite{gerlich92} can be written in polar coordinates as:
\begin{equation}
 \phi_{RF}(r,\theta,t) = V_{RF}\cos(\Omega t)\left(\frac{r}{r_{0}}\right)^{k} \cos(k\theta)
\end{equation}
where $V_{RF}$ is the amplitude of the RF potential applied to the rods, oscillating at frequency $\Omega/2\pi$, and $r_{0}$ is the inner radius of the trap. In the adiabatic approximation where the motion induced by the spatial envelop of the electric field is far slower than the one driven by the RF voltage, the confinement and motion of the ions can be understood as if they were trapped in a static potential called the pseudo-potential which takes a cylindrical symmetric form
\begin{equation}
\phi_{ps}(r) = \frac{Q}{m}\frac{k^2 V^{2}_{RF} }{4 r_{0}^{2} \Omega^{2} } \left(\frac{r^{2}}{r^{2}_{0}}\right)^{k-1}
\end{equation}
with $Q$ and $m$ the electric charge and the mass of the trapped ions. Confinement along the trap axis ($z$-axis) is reached by dc potentials applied to extra electrodes at the end of the trap or outer segments of the rods. It can be approximated by its lowest order expansion
\begin{equation}
\psi_{dc}(r,z) = \frac{1}{2}\frac{m}{Q}\omega_{z}^{2} \left(z^{2} - \frac{r^{2}}{2}\right).
\label{eq:dc_part}
\end{equation}
For a multipole trap with $k>2$, the non-confining radial contribution of $\psi_{dc}(r,z)$ makes the centre of the trap an unstable equilibrium position. It sums up with the confining radial pseudo-potential to form a trapping potential with the profile of a Mexican hat where the minimum of potential is shifted from $r=0$ to a ring of radius $r_{min}$ \cite{champenois10}:
\begin{align}\label{eq:rmin}
  r_{min}^{2k-4} = \frac{1}{k-1}\left(\frac{m\Omega \omega_{z}r_{0}^{k}}{kQV_{RF}}\right)^{2}.
\end{align}
As the Coulomb repulsion also drives the ions away from the trap centre, their radial equilibrium position depends on their number and the strength of the dc potential. Ions settle at $r_{min}$ only if the Coulomb repulsion can be neglected compared to the dc potential \cite{champenois10,marciante12,cartarius13}. Nevertheless, when cold enough, ions are expected to form a hollow core structure which can be described as a ring \cite{champenois10,cartarius13}, a monolayer tube \cite{marciante12}, or a multilayer tube \cite{calvo09}. This organisation can be understood in the mean-field limit as the condition for a Poisson-Boltzmann equilibrium at low temperature \cite{champenois09}.

\section{Observation of laser-cooled ions in an octupole trap}\label{manip}
An octupole ($k=4$) trap \cite{champenois13bis} has been designed to study the equilibrium and dynamical properties of large samples of laser cooled Ca$^+$ ions ($m=40$~a.m.u.). Its inner radius $r_0$ is 3.83~mm and its internal axial length $2z_0$ is 96~mm. The radius $r_d$ of each rod was chosen to obey the relation $r_{d} = r_{0} / (k-1)$ \cite{gerlich92}. The parameters of the trap used throughout this article are $\Omega = 2\pi \times 3.3$~MHz and $V_{RF} = 300$~V if not otherwise indicated.
The design of the trap includes an in-line quadrupole trap where ions are first created by photo-ionisation and laser-cooled before being shuttled to the octupole trap \cite{kamsap15t, kamsap15a}. This strategy avoids contamination of the trap rods by neutral calcium which can be responsible for contact potentials which deform the trapping potential.

The shuttling of the ions to the octupole uses the same method as the one experimentally demonstrated in \cite{kamsap15a} by efficiently achieving transport of large ion clouds between two consecutive quadrupole traps, based on the protocol developed in \cite{pedregosa15}. This transport protocol depends only on the axial profile of the potential and not on the quadrupole nature of the radial trapping, and could thus be extrapolated to transport to a  trap of different geometry, as both traps are operated in a symmetric configuration ($\pm V_{RF} \cos(\Omega t)$) with no grounded electrodes. This assures that the contribution of the RF-potential is null along the symmetry axis. The main difference between the two kinds of traps concerning the trapping along the axis is the screening of the DC voltages by the RF electrodes, which increases with the number of rods, as shown in figure 3 of \cite{champenois13bis}.

Within the octupole trap, ions are observed through the laser-induced fluorescence on the resonance line at 397~nm, driven by the Doppler-cooling laser. The fluorescence is collected by a complex lens of 66.8~mm focal length along a direction perpendicular to the trap $z$-axis and is split between a photomultiplier and an intensified charge-coupled device camera. An image is formed on the camera at a distance of about 60~cm from the lens.

\begin{figure}
    \center
    \scalebox{0.47}{\includegraphics{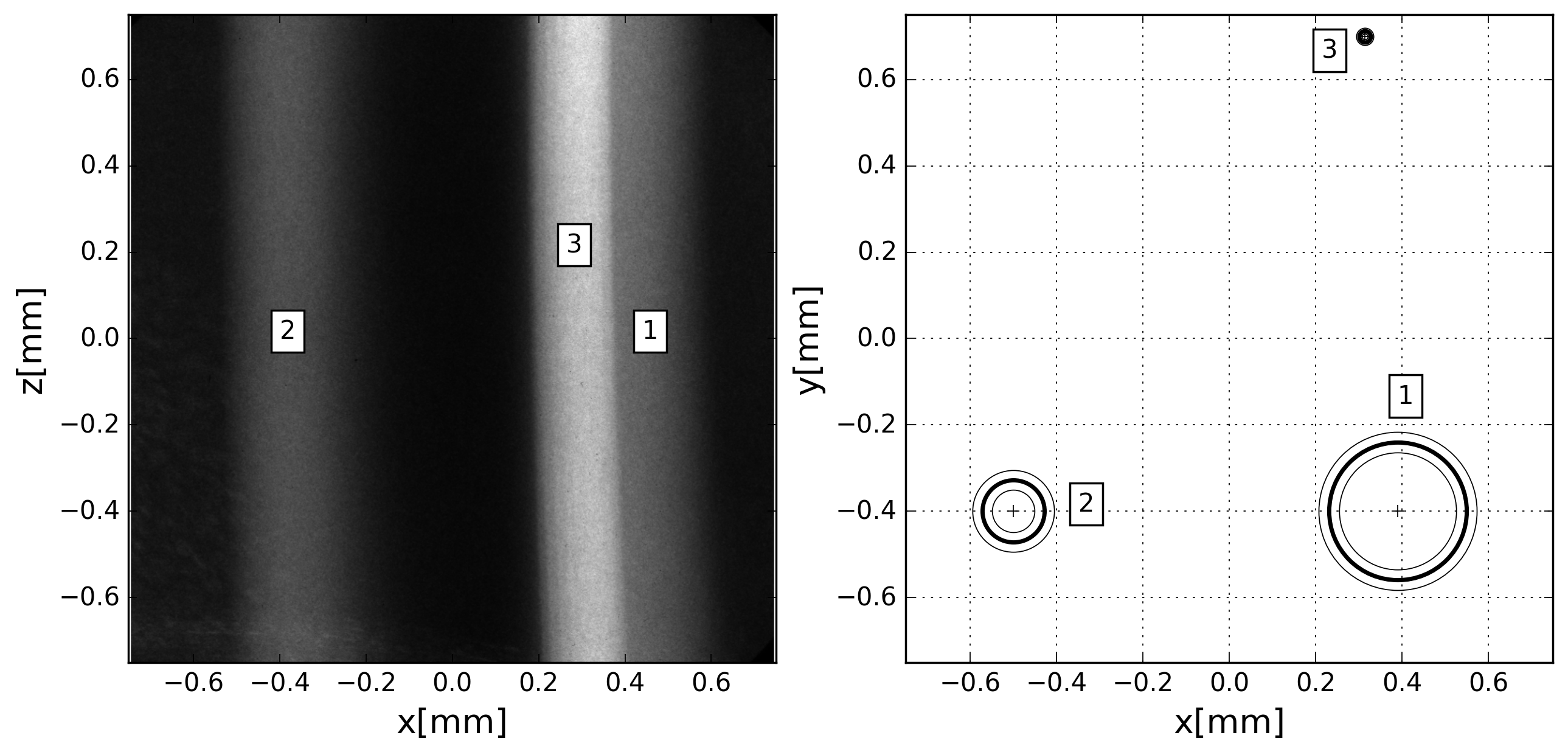}}
    \caption{Left: Photo of the fluorescence of laser-cooled ions trapped in a linear octupole trap, collected in a direction perpendicular to the trap $z$-axis, which is along the vertical direction on the picture. The indicated zero of the axis refers to the centre of the picture and not to the centre of the trap. The focus is chosen such as to give a sharp image of the cloud labelled "3". The applied RF voltage was $V_{RF}=425~V$ and the $\omega_z/2\pi = 3.25kHz$. Right: Representation in the $x$-$y$-plane of the position of the three observed ion clouds with their mean relative ion number scaled by the radius of the circle (thick lines). The thin line radius stands for one standard deviation.
    \label{fig:experimental_crystals}}
\end{figure}

Fig.~\ref{fig:experimental_crystals} shows the fluorescence of the laser-cooled ions in different image depths, three independent clouds rather than a hollow structure can be identified on the picture. In order to increase the number of ions in the octupole region, the accumulation technique described in \cite{kamsap15a} has been used. None of the observed ion clouds is located at the expected radial centre of the trap.
The translation of the collecting optics along the trap axis shows that the three ion clouds are not connected at their ends and the observation of these clouds during the accumulation process reveals that they do not grow identically: during the increase of the number of ions, the cloud which is the largest on Fig.~\ref{fig:experimental_crystals} (labelled "1") is the first one to grow, then followed by "2", the second largest cloud, and "3", the smallest one. When crystallised, each cloud organises like in a very long quadrupole trap, from a long ion chain, containing up to 155 ions \cite{kamsap17} to elongated clouds showing several layers. The ion clouds "1" and "2" are separated by $890\pm5 \mu m$ and "1" and "3" are separated by $1103\pm11 \mu m$. In the plane perpendicular to the axis of the imaging optics, we deduce these distances from the size of the image and the 8.1 magnification of our optics. Along the observation axis, the optical depth is very small ($4\mu m$). We deduce the distance between the clouds from the micrometric displacement of the lens between the positions corresponding to distinct images of the different clouds.

These observations can be explained by the non-degeneracy of the zeroes of the rf electric field because of symmetry breaking. Whatever is the number of electrodes or the spatial distribution of the RF-electric field, the general form of the pseudo-potential is proportional to $|E_{rf}|^2$ where $|E_{rf}|$ is the norm of the amplitude of the total RF electric field created by the whole set of electrodes. In a linear multipole with $2k$ rods, the rf potential in the $(x,y)$ plane is a $(x,y)$ polynomial of order $k$ and each projection of the electric field is then a $(x,y)$ polynomial of order $(k-1)$ defined by $k$ constant parameters. If the geometry is perfect and the $k$ mirror symmetries are fulfilled, the $k$ parameters are constrained and there is only one zero of order $(k-1)$ for the electric field, and it is located at the centre of the trap. If the symmetries are broken, the degeneracy is lifted which gives rise to $(k-1)$ different positions where the RF electric field, and so the pseudo-potential, vanishes. This behaviour has already been identified in a 22-pole trap where 10 isolated ion clouds have been observed \cite{otto09}. The lowest order of the polynomial development of the electric field around each zero is a priori linear, as no symmetry condition can cancel these terms. As a consequence, the pseudo-potential has a local harmonic behaviour around each zero of the electric field, which turns to be the minima of the pseudo-potential.

\begin{figure}
    \center
    \scalebox{0.35}{\includegraphics{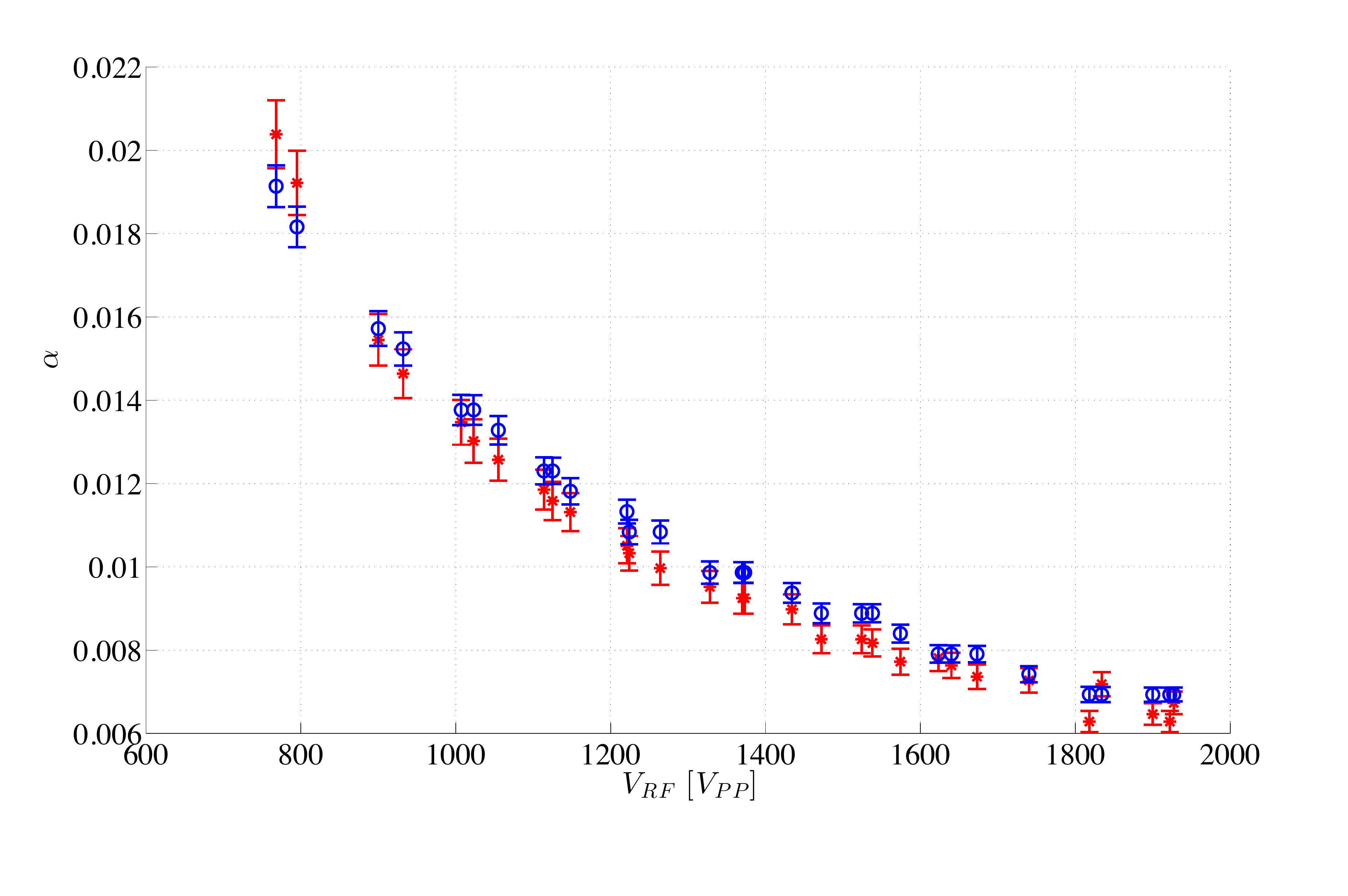}}
    \caption{Aspect ratio $\alpha$ of an ion cloud in the octupole trap (potential well "1") for different trapping voltages $V_{RF}$. Red full circles show measured values, while blue open circles have been calculated for a quadrupole potential. }
    \label{fig:aspectratio}
\end{figure}

To validate the assumption of the harmonic nature of the local potential minima, the parametric excitation of the ion cloud in its gas phase   has been used \cite{champenois01}. By adding an oscillating perturbation to the dc potential, and repeating the experiment for different trapping conditions, we are able to identify resonances with frequencies scaling like $V_{RF}$ and frequencies scaling like $\sqrt{V_{dc}}$, independent of the amplitude of excitation. These scaling laws allow the identification of the eigen-frequencies of motion in the radial pseudo-potential $\omega_x$ and in the axial dc potential $\omega_z$. To check the consistency of the local harmonic potential approach, we measure the aspect ratio $\alpha= R/L$ of the ellipsoid shape formed by an ion cloud that is laser-cooled to the liquid phase in the main local potential well "1". Figure~\ref{fig:aspectratio} shows these values together  with the ones calculated from the aspect ratio  of the trapping potential \cite{turner87, hornekaer02} by using the previously measured motional frequencies $\omega_x$ and $\omega_z$. The comparison of these aspect ratios for different RF voltages shows excellent agreement, and along with the independence of the frequency values from initial conditions, confirms our description as three quadrupole-like potential wells.

The radial sizes of the three clouds (see Fig.~\ref{fig:experimental_crystals}) are different and lead to the assumption that the three local potential minima are not equally populated. The ion number in each cloud can be estimated from the volume of the ellipsoid shaped cloud and its density, once the clouds are cooled to the liquid phase. The radial size is measured from the pictures, the length is deduced from the calculated aspect ratio of the ellipsoid and the density of the ion cloud is calculated from the low temperature limit of the mean field description of the cloud: $n_c=2m \epsilon_0 \omega_x^2/Q^2$ \cite{dubin99}. We assume that the density in the cold limit is the same for the three clouds,  and we find that the repartition does not depend on the total number of trapped ions. Total ion numbers range from $1.1\times 10^4$ to $2.1\times 10^5$. On average, cloud "1" gathers 65\% of the ions with a standard deviation of 10\%, cloud "2" $(29 \pm 9)$\%  and cloud "3" $(6 \pm 3)$\% .

\section{Characterisation of the local minima by numerical simulation}
In order to better understand the behaviour of the described local minima in the octupole trap, we have developed numerical simulations where a controlled error can be introduced. Some similar work was performed in the case of 22-pole traps in \cite{otto09} in the particular case of an angular mispositioning of the rods, due to specificities of the experimental set-up. We have implemented a more systematic approach, and are  interested in the effect on the potential landscape due to the displacement of a single rod.

With our trap geometry, additional potential minima are already observed if an ideal trap geometry is simulated by the widely-used SIMION software \cite{simion}, due to rounding errors associated with the numerical algorithm itself as an increase of the spatial resolution used does not modify the results. To avoid these issues, we use the commercial software CPO \cite{CPO}, which computes the electric field $\vec{E}$ generated by a given electrode geometry based on the Boundary Element Method (BEM). 

A 2D description is considered, which is a rigorous description of a trap with infinitely long and parallel rods. In any linear trap, a misalignment along the $z$-axis can cause an extra variation but the proposed 2D analysis is still relevant as it can be considered as a local description of the radial potential. The positions of the pseudo-potential minima are found by searching for local extrema in $|\vec{E}|^{2}$. Even if the simulations are carried out for a given trap radius $r_0$, the results are independent of this scaling radius if they are expressed in reduced length scaled  units with respect to $r_0$. We write  these rescaled sizes as $\bar{x} = x / r_{0}$.

\subsection{Effect of the displacement of a single rod}
\begin{figure}
	\center
	\scalebox{0.25}{\includegraphics{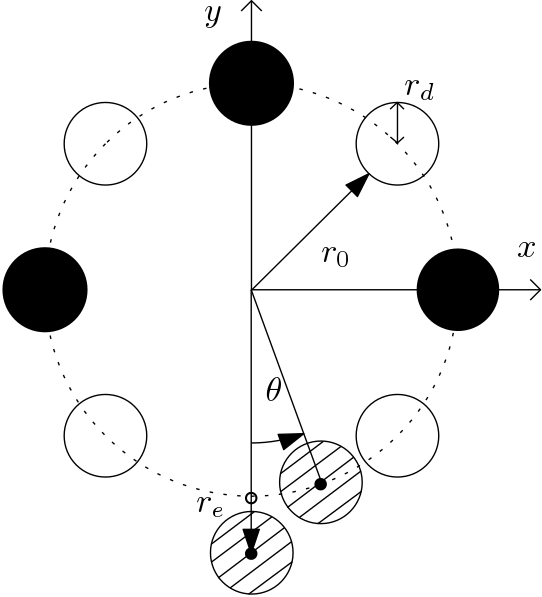}}
	\caption{Scheme in the radial plane of the electrodes of a linear octupole trap. Black and white rods are polarised by $\pm V_{RF}$ voltages. The dash-filled circle corresponds to the displaced rod and the relevant radii and angle are drawn.}
	\label{fig:octo_rods}
\end{figure}

For a systematic study, we assume that all the rods forming the trap are in their nominal position except for a single rod. Its displacement can be a radial or an angular translation from its nominal position or both. The choice of the shifted rod and the definition of the relevant parameters used to describe the geometry are explained in Figure~\ref{fig:octo_rods}.

\begin{figure}
\center
\scalebox{0.4}{\includegraphics{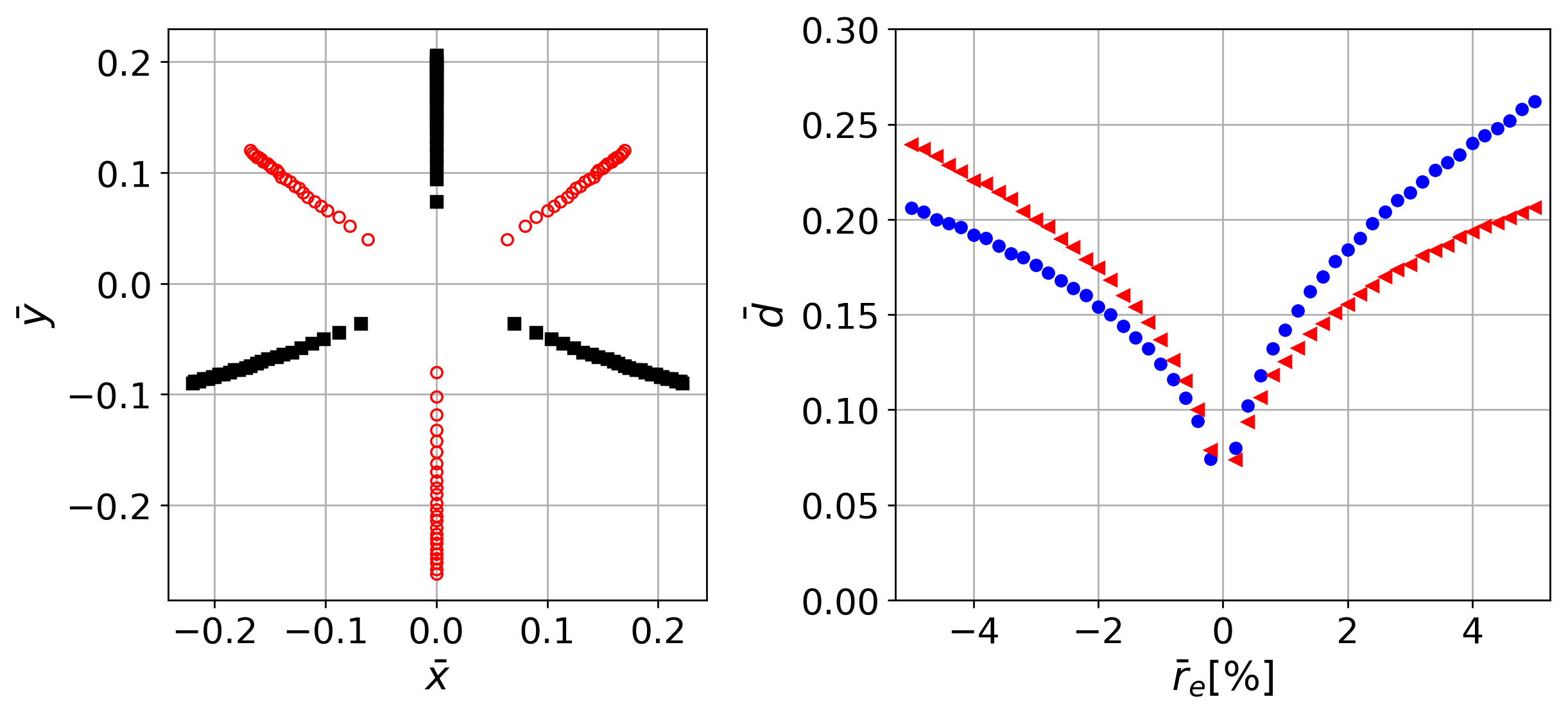}}
\caption{Left: Position of the local potential minima for displacement $\bar{r}_e$ ranging from  0.2 to 5\% of the rod located on the negative $y$-axis. The smallest displacement gives rise to the closest minima to the trap centre. The filled squares correspond to a rod displaced towards the trap centre, the open circles to a rod displaced away from the centre. Right: Distance of the three minima to the trap centre as a function of the normalized rod displacement. The blue circles correspond to the minimum on the $y$-axis and the red triangles to the other two minima. As a convention, $\bar{r}_{e}>0$ stands for a rod at a radius larger than $r_0$, $\bar{r}_{e}<0$ for a rod closer to the centre.}
\label{fig:single_rods_effects_r_traj}
\end{figure}

As a starting point, the effect on the potential landscape due to a purely radial displacement, denoted as $r_{e}$, is explored. While the commercial software CPO uses the BEM method, describing only the surface of the electrodes, the output consist on the electric field on a 2D grid. For the present work, such grid had a resolution of $dx = dy = r_{0}/500$. Figure~\ref{fig:single_rods_effects_r_traj} shows the positions of the three potential minima in the radial plane of an octupole trap, while the rod relative displacement is varied in steps of 0.002. The arrangement of the local minima, a symmetric triangle with respect to the y-axis, reflects the remaining symmetry of the geometry once the rod displacement is introduced. The  triangle is flipped when the displaced rod is moved towards or away from the trap centre.
The distance of each minimum to the trap centre as a function of the rod relative displacement is also plotted in Figure~\ref{fig:single_rods_effects_r_traj}. The singularity at $r_{e} = 0$, means that even very small errors have a significant  impact in the minima's distance to the centre. In the limit case of $r_{e} \to \infty$, the position of the minimum on the $y$-axis moves to $\infty$ and the two remaining symmetric minima are located at $(\bar{x}, \bar{y}) = (\pm 0.312, 0.250)$. \\

A full characterisation of the potential landscape requires the determination of the potential depth associated to each minima. To obtain general results across different trap dimensions, the multipole order scaling factor and the radial size of the trap have been introduced:

\begin{align}\label{eq:pseudo}
  \phi_{ps}(r) = \frac{k^{2} Q V_{RF}^{2} }{4 m r_{0}^{2} \Omega^{2} } \times \left|\frac{r_{0} \vec{E}}{k V_{RF}}\right|^{2}
\end{align}

The right term has no dimension and allows to compare different traps. In the following, we plot this reduced potential and write it as $\bar{\phi}_{ps} =\left|\frac{r_{0} \vec{E}}{k V_{RF}}\right|^{2}$.

\begin{figure}
	\center
	\scalebox{0.4}{\includegraphics{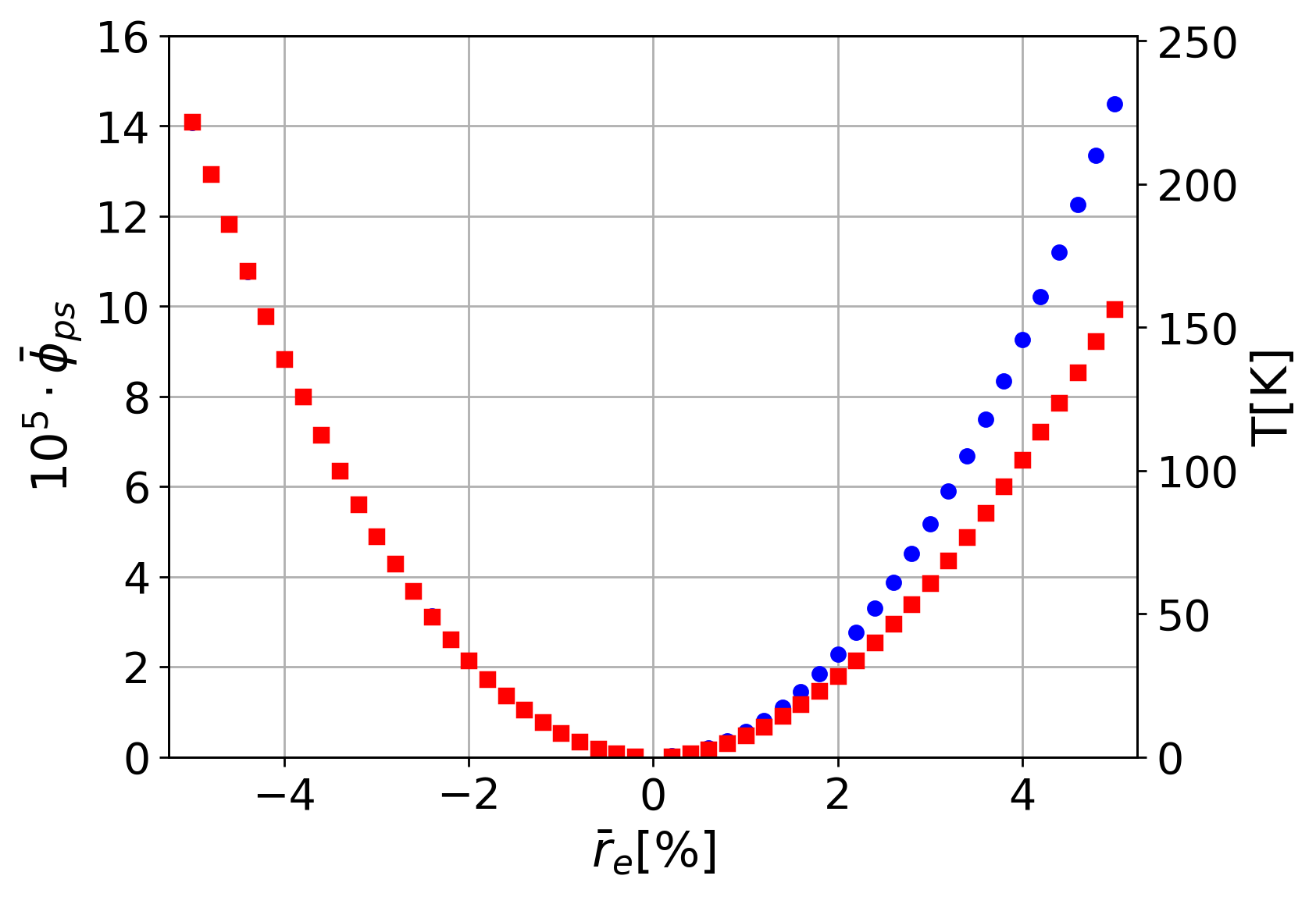}}
	\caption{ Depth of the three local potential minima versus a single rod displacement (same sign convention as in Fig.~\ref{fig:single_rods_effects_r_traj}). The blue circles correspond to the minimum aligned with the displaced rod and the centre of the trap, the red squares to the two other minima. For $\bar{r}_e<0$, values are too close to be distinguished. Two scales for the pseudo-potential are used. Left $y$-axis: $\bar{\phi}_{ps}$. Right $y$-axis: $Q \phi_{ps}/k_B$ (K) , the scaling factor is computed with our experimental parameters (Ca$^+$, $r_{0}=3.83$~mm, $\Omega/2\pi = 3.3$~MHz and $V_{RF}=300$~V). }
	\label{fig:single_rods_effects_r_depth}
\end{figure}

For the definition of the potential depth from the 2D pseudo-potential grid, we have used the difference between the potential at the minimum position and the potential at the closest saddle point. The evolution of the three very similar potential depths is shown in figure~\ref{fig:single_rods_effects_r_depth} where a quadratic variation with $\bar{r}_{e}$ is clearly visible. 

Figure~\ref{fig:single_rods_effects_r_depth} also shows the potential depths for the experimental parameters of our trap, if the influence of the axial confinement is not taken into account. These potential depths are lower than 5~K for a rod mis-positioning of $30~\mu$m and reach 200~K when the radial shift is as large as $190~\mu$m.

We now move to the case where a single rod is displaced by an angular shift (see Fig~\ref{fig:octo_rods}). To compare to the case of a radial displacement, we use the effective angular displacement expressed as 
\begin{equation}
\theta = r_e/(r_0+r_d)= 0.74 \;\bar{r}_e
\end{equation}
where the relationship $r_{d} = 0.355 r_{0}$ \cite{ramarao00} has been used.
Due to the symmetry of the problem, only positive displacements have been studied (see Fig.~\ref{fig:octo_rods} for notations).

In the case of an angular displacement, there is no preserved symmetry but the three minima form a triangular configuration very similar to the previous case of a purely radial displacement. 
Comparing the relative distances from these minima to the centre of the trap, we observe that an angular misalignment leads to distances roughly half of those observed for equivalent radial displacement. The potential depths of the three local minima  are very close and show a larger discrepancy with the previous case: for the same relative rod displacement, the potential depth for an angular displacement is smaller by a factor of the order of 35 compared to the one induced by a radial displacement.

\subsection{Random displacement of all rods}
The displacement  of the single  rod  presented above gives insight into the effect of the radial and angular misalignment and shows that a radial misalignment generates  deeper potential minima than an angular one. However, a realistic situation will present positioning errors in all but one electrodes. In the following, we analyse a geometric configuration where each rod centre is shifted from its nominal position by the same distance $r_{e}$, but in a random direction within the $2\pi$ allowed range.

\begin{figure}
\center
\scalebox{0.5}{\includegraphics{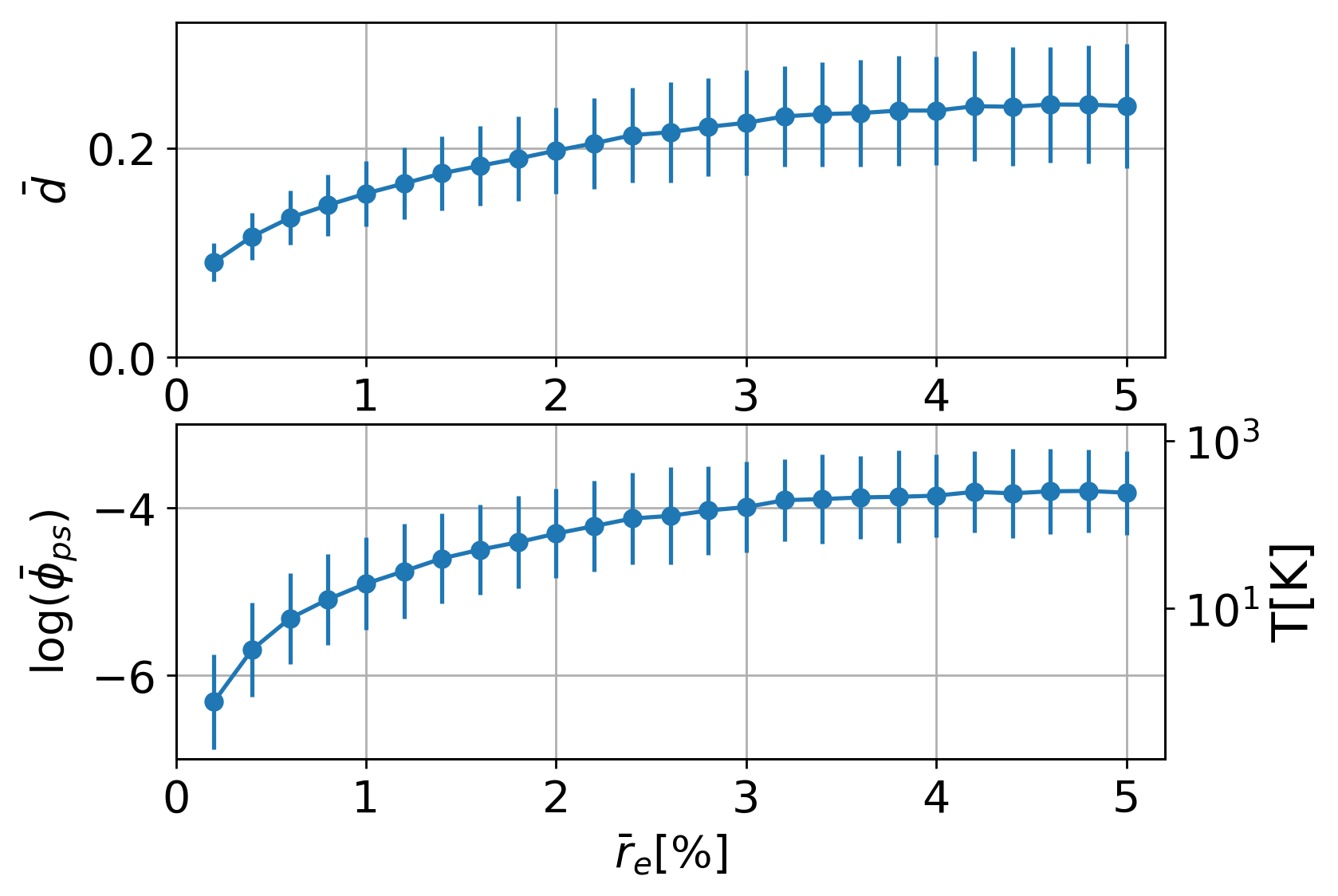}}
\caption{Random radial variation of all trap rods. Top: Mean value of the distance of the minima to the trap centre versus the relative distance of the rods to their nominal position $r_{e}/r_{0}$. Bottom, left scale : mean  of normalised potential depth $\log{\bar{\phi}_{ps}}$ versus $r_{e}/r_{0}$. Bottom, right scale : mean potential depth in temperature units $Q\phi_{ps}/k_B$ (K), the scaling factor is computed with our experimental parameters (Ca$^+$, $r_{0}=3.83$~mm, $\Omega/2\pi = 3.3$~MHz and $V_{RF}=300$~V).}
\label{fig:avg_stats_several_re}
\end{figure}

For 1000 random sets of rod position with a chosen relative distance $r_{e}/r_{0}$, we compute the average distance, $\bar{d}$, and its standard deviation, $\sigma_{d}$, of the local potential minima to the trap centre. Repeating the statistics for different values of $r_{e}/r_{0}$ leads to the top graph on Figure~\ref{fig:avg_stats_several_re} where the mean values of $\bar{d}$ are shown. The error bars correspond to one standard deviation.

Similarly, an evolution of the potential depth with respect to $r_{e}/r_{0}$ leads to the bottom graph on Figure~\ref{fig:avg_stats_several_re}. Due to the dispersion of values over several orders of magnitude, the statistics have been performed over the $log(\bar{\phi}_{ps})$ values. Although the data statistics present a large scattering around their mean values, these general results can be used for estimating the mechanical error that can be tolerated for a given trap size and a given application, identified here by the temperature of the trapped sample. For example, using the experimentally measured positions of the potential minima in our set-up, their average distance to the trap centre is $0.17r_0$. In Figure~\ref{fig:avg_stats_several_re}, this value of $\bar{d}$ corresponds to $\bar{r}_{e} = 1.28 \%$, and therefore to an average potential depth of 32~K, which agrees qualitatively with our experiments.

\section{Ion number distribution in additional local minima}
\begin{figure}
	\center
	\scalebox{0.5}{\includegraphics{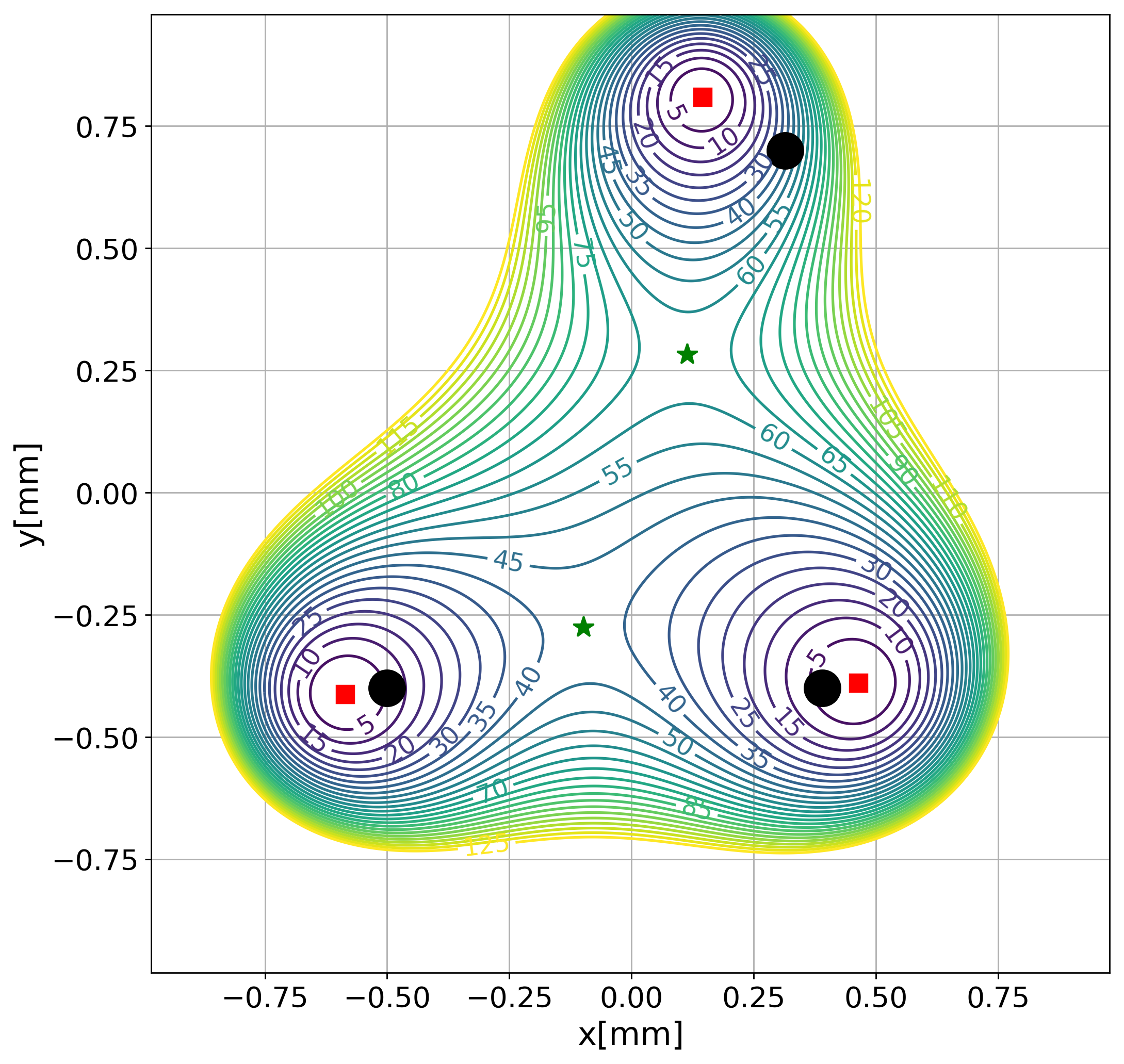}}
	\caption{Pseudo-potential given by CPO for the best agreement with the experimentally determined local minima (black circles) out of the 2000 test cases. The red squares represent the minima of the best agreement, and the green stars the saddle points used to compute the potential. The numerical values in the iso-lines correspond to a temperature in Kelvin when using parameters of our experimental set-up. See text for details.}
	\label{fig:min_pos_exp_vs_cpo}
\end{figure}

In the previous section, we have proposed a correlation between the mean positioning error of each rod and the average potential depth and position of the local potential  minima. Yet, this geometric description does not explain the observed unbalanced ion distribution among the three local minima.
In order to achieve a better understanding of the experimental set-up and of the role of the potential local minima in the ion organisation, we have inverted the approach to numerically find the rod positions that lead to an ion distribution similar to the one that has been experimentally observed. In principle, an electrode distribution that reproduces the measured minima positions could be found by implementing a non-linear algorithm that minimises the difference between the experimentally observed and numerically simulated positions of the potential minima. While this is indeed possible, the large number of parameters (14) that are required to define the $(x,y)$ coordinates of seven of the trap rods, makes the task daunting and has not been undertaken. We have rather chosen a method that consists in generating random sets of positions of the rods, where each rod can have its distance to its nominal value ranging from 0 to $0.1 r_0$. For each geometric configuration, the pseudo-potential is computed to find the positions of the three potential minima, and the average smallest mean difference with the experimental position of the minima is calculated. Figure~\ref{fig:min_pos_exp_vs_cpo} shows the best agreement out of 2000 random position distributions. In this case, the average distance between the potential minima and the trap centre is $0.17r_0$, for a mean displacement of the rods of $0.016r_0$ which corresponds to $60~\mu$m in our set-up. The potential depth of the numerically obtained minima is 42~K for the two minima which are aligned along the $x$-axis and 62~K for the one with a larger $y$-value. These values, 42~K and 62~K, are in a good qualitative agreement with the value obtained from the Figure~\ref{fig:avg_stats_several_re}: 32~K.

The obtained relative values for the potential depths of the three local minima are clearly not sufficient to explain the discrepancy between the ion distribution observed experimentally.

In order to verify that the methodology put forward is valid, the complete potential landscape shown in Fig.~\ref{fig:min_pos_exp_vs_cpo} has been used for a molecular dynamics simulation of the trapped ions. The simulation used follows the approach described in \cite{marciante10}. To reach a 3D confinement, in addition to the computed RF-potential of the octupole trap, we add a static axial potential of $\omega_{z}/2\pi = 15$~kHz, which is the typical value we can reach in our very long trap. The numerical code takes into account the force induced by the RF electric field (no pseudo-potential approach), and implements a laser-cooling model which takes into account the photon recoil on absorption and emission (see \cite{marciante10} for details of the laser-cooling algorithm). The cooling laser beam propagates along the symmetry axis of the trap and thus does not induce a symmetry breaking by the radiation pressure.

Even if the number of simulated ions is smaller than the number of ions in the experiments, the simulation reproduces remarkably well the measured repartition for two different sizes of the sample : for N = 1536 ions the repartition is 68\%, 28\%, 4\% and for N=3072, we find a similar ion distribution: 64\%, 32\%, 4\%, compared to the experimental averages 65\%, 29\%, 6\% (see section~\ref{manip}). As the experimental results do not show any number dependence in the repartition of the ions among the three minima, we interpret the very good agreement between the measured and calculated ion distribution as a confirmation of the relevance of our description of the geometric misalignment of the trap.

\section{Conclusion}
We have presented experimental evidences of symmetry breaking in a macroscopic RF octupole trap. Three potential minima are observed, which correspond to the $k-1$ roots of the RF-electric field in the radial plane for a non-symmetric case. These additional potential minima are locally very well described by a quadrupole trap behaviour and are not equally populated. The existence of such minima in addition to the multipole potential might be responsible for the temperature limits in buffer gas cooling which have been recently discussed \cite{endres17}.

Our numerical simulations show that the slightest asymmetry in the radial mechanical mounting of the trap is sufficient to generate additional potential wells with depths that  trap laser-cooled ions. We have therefore started to tackle the in-situ correction of the multipole potential, based on the numerical method presented above, and will present this work in a future article.
 
Full control of the minima positions allows to trap parallel ion strings with a tunable interaction, as discussed in \cite{marciante11}, and opens the way for novel entanglement protocols.

\section*{Acknowledgements}
This work has been financially supported by ANR (ANR-08-JCJC-0053-01),
CNES (contracts no. 116279 and 151084), and R\'egion PACA. M.R.K.
acknowledges financial support from CNES and R\'egion
Provence-Alpes-C\^ote d'Azur.

\bibliographystyle{tfp}

\end{document}